\documentclass[letterpaper, conference, 10pt]{ieeeconf}

\usepackage{multirow}
\usepackage{amsmath}

\usepackage{epsfig}
\usepackage{hyperref}
\usepackage[dvipsnames]{xcolor}

\IEEEoverridecommandlockouts
\begin{document}

\title{\LARGE \bf Automatic Classification of Knee Rehabilitation Exercises
Using a Single Inertial Sensor: a Case Study}

\author{
Antonio Bevilacqua$^{1*}$,
Bingquan Huang$^{1}$,
Rob Argent$^{1, 2}$,
Brian Caulfield$^{1}$,
Tahar Kechadi$^{1}$
\thanks{$^{*}$Correspondence: \tt\footnotesize antonio-bevilacqua@insight-centre.org}
\thanks{$^{1}$Insight Centre for Data Analytics, University College of Dublin,
Ireland.}
\thanks{$^{2}$Beacon Hospital, Dublin, Ireland}}

\maketitle

\thispagestyle{empty}
\pagestyle{empty}

\begin{abstract}
  Inertial measurement units have the ability to accurately record the
  acceleration and angular velocity of human limb segments during discrete joint
  movements. These movements are commonly used in exercise rehabilitation
  programmes following orthopaedic surgery such as total knee replacement. This
  provides the potential for a biofeedback system with data mining technique for
  patients undertaking exercises at home without physician supervision. We
  propose to use machine learning techniques to automatically analyse inertial
  measurement unit data collected during these exercises, and then assess
  whether each repetition of the exercise was executed correctly or not. Our
  approach consists of two main phases: signal segmentation, and segment
  classification. Accurate pre-processing and feature extraction are paramount
  topics in order for the technique to work. In this paper, we present a
  classification method for unsupervised rehabilitation exercises, based on a
  segmentation process that extracts repetitions from a longer signal
  activity. The results obtained from experimental datasets of both clinical and
  healthy subjects, for a set of 4 knee exercises commonly used in
  rehabilitation, are very promising.
\end{abstract}

\section{Introduction}

With the ever-increasing demand for more efficient healthcare delivery,
home-based exercise rehabilitation forms the mainstay of rehabilitation after
injury or operation \cite{AAOS}. This places an increasing emphasis on the
patient’s own self-management skills to maximise the outcome of surgery, yet
many fail to adhere to their prescribed exercise programme \cite{Bassett03}.
Even for those that do adhere to their programme, confusion with the exercise
technique and not remembering to complete the programme are common problems
\cite{Smith05}. Errors in exercise technique due to insufficient range of
movements, time under muscle tension, or biomechanical alignment, have the
potential to detract from the outcome of surgery. With assessment of technique
typically taking place in the clinic, there is the potential for several weeks
of sub-optimal performance between visits. Therefore, a biofeedback system that
can be used in the home has the potential to improve exercise performance
technique, and hence maximise the outcome of rehabilitation after surgery.  With
the use of inertial measurement units (IMUs), capable of sampling physical
motion characteristics, such as acceleration and angular velocity, exercises can
be accurately evaluated using machine learning classification techniques
\cite{Chen_2015, BingPaper, BrianJournal}. In this paper, we present the
implementation of an automatic exercise classification workflow, illustrating
the general system architecture and detailing the phases related to data
collection and cleaning, segmentation, feature extraction, and
classification. The proposed architecture is integrated into an Android
application patients can use to receive real-time biofeedback for their exercise
performance. This paper aims to describe in detail the methodological approach
used during the phases of data composition, preparation, segmentation and
classification of this Android biofeedback application. It will also present and
discuss experimental results and highlight future extensions to this work.

\section{Related work}
\label{section:relatedWorks}

Current biofeedback methods for biomechanical analysis consist of force plates,
camera-based motion capture systems such as Microsoft Kinect, and IMUs
\cite{Giggins2013}. Biofeedback systems using IMUs have been investigated in
numerous populations including falls, neurological rehabilitation, physical
activity and exercise rehabilitation \cite{Giggins2013}. Yet examples of these
systems in clinical practice in the physical therapy field at present do not
classify exercise technique, rather guide the user through the exercise
programme by tracking repetitions, with interactive educational features along
the way \cite{Smittenaar2017}.  Chen \textit{et al} \cite{Chen_2015} assessed
the performance of a classification system for knee rehabilitation exercises
based on data collected from 3 IMU sensors. They leveraged on the shank angle
variation to segment the exercise signal, and classified the exercises using a
mixture of time-domain and frequency-domain features, along with specific
information about the angle variation. Their system detects multiple exercise
deviations for 3 different knee rehabilitation exercises.

Previous work within this research group has identified the ability to classify
exercise performance in commonly prescribed exercises following total knee
replacement, and that by reducing the number of sensors to a single IMU,
satisfactory levels of accuracy are maintained \cite{Giggins2013}.  In this
paper, we extend the work from Bingquan \textit{et al} \cite{BingPaper}, who
determined that binary classification for knee exercises has higher accuracy
scores compared to the corresponding multilabel classification, and that the
shin resulted to be the best sensor location for most of the target
exercises. In their work, they evaluated the use of multiple sensors to improve
segmentation and classification accuracy. However, as we intend to test our
system in the home environment with a patient population, we are only using a
single sensor located on the shin. We aim to highlight the most effective
classification method that can be used for each exercise, and the accuracy these
methods deliver. We also outline the method of segmentation for individual
repetitions and the results achieved with this technique.

\section{Data and Methodology}
\label{section:methodology}
The aim of this study is to assess the performance of a single IMU
classification system for four different knee rehabilitation exercises. The
position of the unit on the shin is shown to be optimal in terms of detecting
deviations that may happen during the exercise execution \cite{Giggins}, and is
also convenient for easy placement by the patient. The sensor is placed in a
neoprene sleeve at the midpoint of the shin, in the midline of the thigh on the
anterior aspect.

\subsection{Exercises and labels}
Four popular rehabilitation exercises for the knee are targeted in this paper:
the heel slide (HS), the seated knee extension (SKE), the inner range quadriceps
(IRQ) and the straight leg raise (SLR). HS, IRQ and SLR require the subject to
be in a lying position, while SKE requires the subject to be in a sitting
position, as extensively described in \cite{Giggins}. Based on the work in
\cite{BrianJournal}, the physical deviations that can be detected with a single
shin sensor for the target exercises are excessive hip external rotation (ER)
for HS, lack of full knee extension (KF) for SKE, excessive hip flexion (HF) for
IRQ and inability to maintain full knee flexion (KF) for SLR.

\subsection{Study Subjects and Dataset Composition}
A balanced dataset of correctly and incorrectly performed exercises is collected
from a mixed group of 44 clinical subjects and 10 healthy subjects. Both groups
performed 10 correct repetitions and 10 incorrect repetitions for each one of
the 4 target exercises. In addition, the healthy subjects performed the
exercises simulating fatigue conditions, with different pausing times between
every pair of consecutive repetitions, or holding times at the isometric peak of
each repetition, and mixing correct and incorrect repetitions. All data
collection was supervised by a Chartered Physiotherapist.

\subsection{Data Collection, Preparation and Preprocessing}
The Shimmer3 IMU device \cite{Burns2010} was used to collect data using two
different kinematic sensors, the triaxial digital accelerometer KXRB5-2042, and
the triaxial digital gyroscope MPU-9150 \cite{ShimmerSensing}. We kept the
device configuration consistent through all the observations, as follows:

\begin{itemize}
\item Sampling rate of 102.4 Hz for both accelerometer and gyroscope (a total of
  1024 samples are collected every 10 seconds).
\item Accelerometer range of $\pm 2g$.
\item Gyroscope range of 500dps.
\end{itemize}

The device was then calibrated using the Shimmer 9 DoF application. Each sensor
axis has a specific baseline and orientation. The baseline represents the value
sampled by the Shimmer when gravity only is applied to the sensor. The
orientation is used to understand whether the sensor is facing up or down. The
task of ensuring that all the units are properly configured, calibrated, and
oriented, is paramount during the process of data collection, as baseline values
and orientation references play an important role during the segmentation phase.

For each performed exercise, six numerical vectors are directly sampled by the
Shimmer. That are, the \textbf{acceleration} and \textbf{angular velocity}
signals over the axes $x$, $y$ and $z$. In addition, three new vectors are
analytically derived and used in the feature extraction phase. They are the
\textbf{magnitude}, the \textbf{pitch}, and the \textbf{roll} vectors. The
magnitude represents the overall sensor velocity, while the pitch and the roll
represent the rotation over the lateral axis and longitudinal axis,
respectively.  The nine raw signal vectors are smoothed by using a
4\textsuperscript{th} order lowpass Butterworth filter, so that the noise
introduced by the elastic vibration of the strap that keeps the Shimmer in place
is removed. A subsequent min-max normalization is applied to all signals.

\subsection{Signal Segmentation}
\begin{figure}[h!]
  \centering
  \includegraphics[width=\columnwidth]{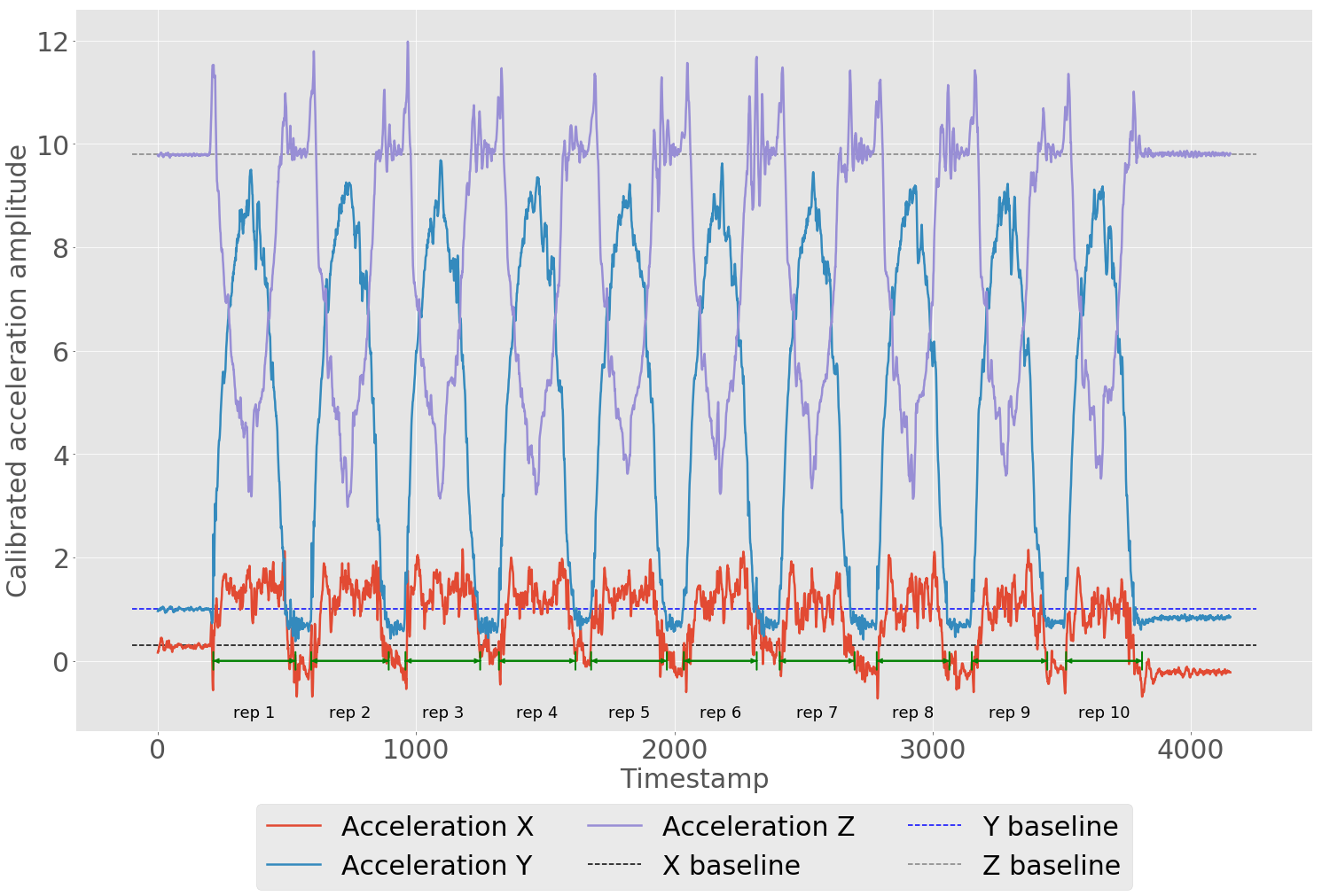}
  \caption{Segment coordinates extracted from an HS exercise.}
  \label{img:segmentedExercise}
\end{figure}

We apply a template matching algorithm to the resulting nine signal
vectors. The segmentation algorithm returns a set of coordinates used to trace
the repetitions within an exercise, as shown in Figure
\ref{img:segmentedExercise}. Our technique exploits the pausing windows
interleaving the exercise repetitions in order to detect silent data within the
signals, i.e., data points that fall into inactivity periods.

Our signal segmentation approach is described in detail in the following steps:

\begin{itemize}
\item Zero velocity points are analytically derived from the signals, by using
  the expected signal baselines as reference values. Several thresholds are used
  to single out and discard particularly noisy points.
\item The obtained zero velocity points are clustered with the k-means
  algorithm. The centroid of each cluster is used as the representative point for
  a signal window that potentially contains silent data only. All the centroids
  compose the set of candidate points.
\item Signal chunks are extracted by using all the pairs of subsequent
  candidate points. For each chunk, a set of features is extracted. These
  features are length, height, standard deviation, skewness, kurtosis, and the
  first 20 coefficients of the signal Fourier transformation.
\item Each chunk is classified with a Hoeffding tree, previously
  trained with a template dataset containing ideal characteristics for properly
  and improperly executed repetitions.
\item All the candidate points that divide the signals into positively
  classified chunks are returned as cutting points.
\end{itemize}

\subsection{Feature Extraction}
Once the exercises are chunked into the composing repetitions, a set of 356
features is extracted from each of them. As previously stated, each repetition
contains 9 different signal vectors. For each vector, two different groups of
features are calculated. Static features are composed of mean, median, standard
deviation, variance, range, kurtosis, skewness, maximum, minimum, positive mean,
negative mean, sum of absolute differences, 1\textsuperscript{st} quartile,
3\textsuperscript{rd} quartile, and the correlation index between pitch and roll
(the last one is only calculated for the pitch and the roll vectors). Dynamic
features are composed of energy, energy ratio, energy average, harmonic ratio,
energy entropy, and the first 20 coefficients of the signal Fourier
transformation.

\subsection{Repetition Classification}
The previously extracted features are used to train a set of classifiers. The
adopted models are logistic regression, support vector machine trained with the
SMO technique, adaptive boosting, random forest, and J48. The WEKA library
\cite{weka} was used for the classification phase. The performance for each
model is assessed with a 5-fold cross-validation process. In order to avoid
overfitting, folds are generated by dividing the datesets over the patients, so
that, for each fold, any given subject is included either in the training set or
in the test set. During the classification process, the correctly segmented
repetitions only are included in the dataset.

\subsection{Real World Usage}
Newly executed exercises are sampled, segmented and classified with a native
Android application. The Shimmer3 Bluetooth capabilities allow the application
to stream the sampled data, collect them, then perform segmentation, feature
extraction and segment classification as described in the above sections.

\begin{figure}[]
  \centering
  \includegraphics[width=\columnwidth]{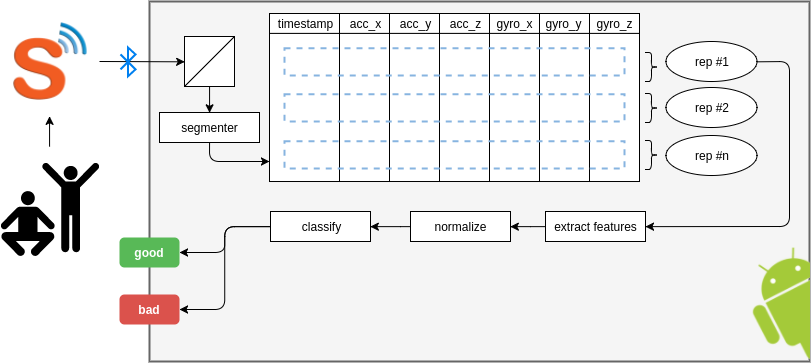}
  \caption{An overview of the classification system}
  \label{img:workflow}
\end{figure}

The workflow, illustrated in Figure \ref{img:workflow}, shows the exercise
supervision platform from a high-level perspective. Previously trained and
serialized classifiers and segmenters are loaded within our Android
application. Each exercise uses the well-suited learning model, according to the
results obtained during the training phase.

\section{Experimental results}
\label{section:experimentalResults}
The first step in the evaluation of our classification procedure is to assess
the performance of the segmentation process, as the quality of the extracted
features highly depends on the cutting points obtained by the segmentation
algorithm. A simple counting of correctly and incorrectly segmented repetitions
is executed. Visual inspection is required in order to validate the obtained
segments, as there is no ground truth based on timestamps that can be used to
determine the goodness of the cutting points. As described by equation
\ref{math:sgmAccuracyFormula}, for each of the $n$ exercises in the dataset the
absolute difference between the actual number of segments $reps_i$ and the
extracted segments $segm_i$ is calculated, then subtracted from the actual
number of segments.

\begin{equation}
  A_{S} = \frac{\sum_{i = 1}^n(reps_i - |reps_i - segm_i|)}{\sum_{i = 1}^nreps_i}
  \label{math:sgmAccuracyFormula}
\end{equation}

Table \ref{table:sgmAccuracyScores} shows the accuracy scores obtained for each
combination of exercise and dataset. It also includes the total number of
performed repetitions for the exercises.

\begin{table}[h!]
  \centering
  \caption{Segmentation accuracy scores}
  \begin{tabular}{|c|| c c c|}
    \hline
    Exercise & Dataset & Repetitions & Accuracy \\
    \hline
    \multirow{3}{*}{HS} & \multicolumn{1}{|c}{lab} & \multicolumn{1}{c}{300} & \multicolumn{1}{c|}{0.97} \\
    & \multicolumn{1}{|c}{clinical} & \multicolumn{1}{c}{714} & \multicolumn{1}{c|}{1} \\
    & \multicolumn{1}{|c}{mixed} & \multicolumn{1}{c}{1014} & \multicolumn{1}{c|}{0.99} \\
    \hline
    \multirow{3}{*}{SKE} & \multicolumn{1}{|c}{lab} & \multicolumn{1}{c}{857} & \multicolumn{1}{c|}{0.98} \\
    & \multicolumn{1}{|c}{clinical} & \multicolumn{1}{c}{754} & \multicolumn{1}{c|}{1} \\
    & \multicolumn{1}{|c}{mixed} & \multicolumn{1}{c}{1611} & \multicolumn{1}{c|}{0.99} \\
    \hline
    \multirow{3}{*}{IRQ} & \multicolumn{1}{|c}{lab} & \multicolumn{1}{c}{695} & \multicolumn{1}{c|}{0.91} \\
    & \multicolumn{1}{|c}{clinical} & \multicolumn{1}{c}{661} & \multicolumn{1}{c|}{0.99} \\
    & \multicolumn{1}{|c}{mixed} & \multicolumn{1}{c}{1356} & \multicolumn{1}{c|}{0.95} \\
    \hline
    \multirow{3}{*}{SLR} & \multicolumn{1}{|c}{lab} & \multicolumn{1}{c}{639} & \multicolumn{1}{c|}{0.97} \\
    & \multicolumn{1}{|c}{clinical} & \multicolumn{1}{c}{658} & \multicolumn{1}{c|}{0.99} \\
    & \multicolumn{1}{|c}{mixed} & \multicolumn{1}{c}{1297} & \multicolumn{1}{c|}{0.98} \\
    \hline
  \end{tabular}
  \label{table:sgmAccuracyScores}
\end{table}

\begin{table*}[!htbp]
  \centering
  \caption{Classification scores}
  \begin{tabular}{|c|c||ccc||ccc||ccc||ccc|}
    \hline
    &                               & \multicolumn{3}{c|}{HS}                                             & \multicolumn{3}{c|}{SKE}                        & \multicolumn{3}{c|}{IRQ}                     & \multicolumn{3}{c|}{SLR}                          \\ \cline{3-14}
    \multirow{-2}{*}{Model}       & \multirow{-2}{*}{Metric (\%)} & lab   & clinical                     & mixed                        & lab   & clinical & mixed                        & lab   & clinical                     & mixed & lab   & clinical   & mixed                        \\ \hline
    & accuracy                      & 78.69 & 91.75                        & 92.61                        & 87.34 & 93.12    & 86.05                        & 73.35 & 84.98                        & 85.33 & 71.79 & 88.93                        & 85.6       \\
    & precision                     & 81.57 & 89.15                        & 91.44                        & 88.78 & 92.93    & 92.56                        & 80.63 & 88.88                        & 86.14 & 76.24 & 88.69                        & 89.12      \\
    \multirow{-3}{*}{Logistic}    & recall                        & 72.97 & 87.89                        & 90.98                        & 86.64 & 92       & 93.75                        & 69.66 & 87.81                        & 82.68 & 83.48 & 83.57      & 88.87                        \\ \hline
    & accuracy                      & 94.39 & 96.9                         & 96.92                        & 95.38 & 94.32    & {\color[HTML]{32CB00} 96.70} & 74.62 & 89.03                        & 85.6  & 72.9  & {\color[HTML]{32CB00} 94.41} & 88.13      \\
    & precision                     & 93.57 & 95.28                        & 95.96                        & 89.11 & 93.32    & {\color[HTML]{32CB00} 96.12} & 78.43 & 88.27                        & 86.75 & 77.92 & {\color[HTML]{32CB00} 92.94} & 91.09      \\
    \multirow{-3}{*}{SMO}         & recall                        & 92.9  & 95.21                        & 95.84                        & 93.32 & 91.75    & {\color[HTML]{32CB00} 96.68} & 65.4  & 86.92                        & 82.47 & 78.97 & 88.47      & {\color[HTML]{32CB00} 89.74} \\ \hline
    & accuracy                      & 92.99 & 94.86                        & 95.52                        & 96.63 & 95.92    & 94.13                        & 70.99 & 84.66                        & 81.14 & 77.86 & 88.98                        & 86.86      \\
    & precision                     & 92.87 & 93.94                        & 94.52                        & 94.05 & 94.97    & 93.82                        & 74.60 & 84.41                        & 81.49 & 80.69 & 89.26                        & 86.24      \\
    \multirow{-3}{*}{Ada Boost}   & recall                        & 92.14 & 93.75                        & 94.23                        & 96.06 & 94.64    & 94.68                        & 53.36 & 80.29                        & 74.55 & 79.88 & 82.24      & 84.94                        \\ \hline
    & accuracy                      & 95.48 & {\color[HTML]{32CB00} 97.72} & 97.57                        & 96.17 & 95.80    & 93.11                        & 73.61 & {\color[HTML]{32CB00} 90.64} & 84.57 & 76.62 & 87.89                        & 84.15      \\
    & precision                     & 96.43 & 97.72                        & {\color[HTML]{32CB00} 97.98} & 91.24 & 95.38    & 94.72                        & 76.14 & {\color[HTML]{32CB00} 90.08} & 84.98 & 76.99 & 90.46                        & 86.06      \\
    \multirow{-3}{*}{Rnd. Forest} & recall                        & 95.84 & 97.75                        & {\color[HTML]{32CB00} 97.84} & 95.17 & 94.96    & 95.74                        & 57.56 & {\color[HTML]{32CB00} 88.97} & 79.67 & 75.54 & 84.07      & 85.28                        \\ \hline
    & accuracy                      & 89.53 & 95.01                        & 94.95                        & 94.44 & 93.05    & 91.75                        & 64.06 & 81.91                        & 77.34 & 76.67 & 86.78                        & 81.44      \\
    & precision                     & 93.72 & 94.6                         & 94.98                        & 92.69 & 95.26    & 93.17                        & 76.22 & 81.09                        & 81.07 & 79.58 & 90.51                        & 86.7       \\
    \multirow{-3}{*}{J48}         & recall                        & 93    & 94.64                        & 94.86                        & 95.02 & 95       & 94.01                        & 64.4  & 78.41                        & 74.6  & 79.35 & 84.13      & 86.29                        \\ \hline
  \end{tabular}
  \label{table:clsScores}
\end{table*}

\begin{figure}[]
  \centering
  \includegraphics[width=\columnwidth]{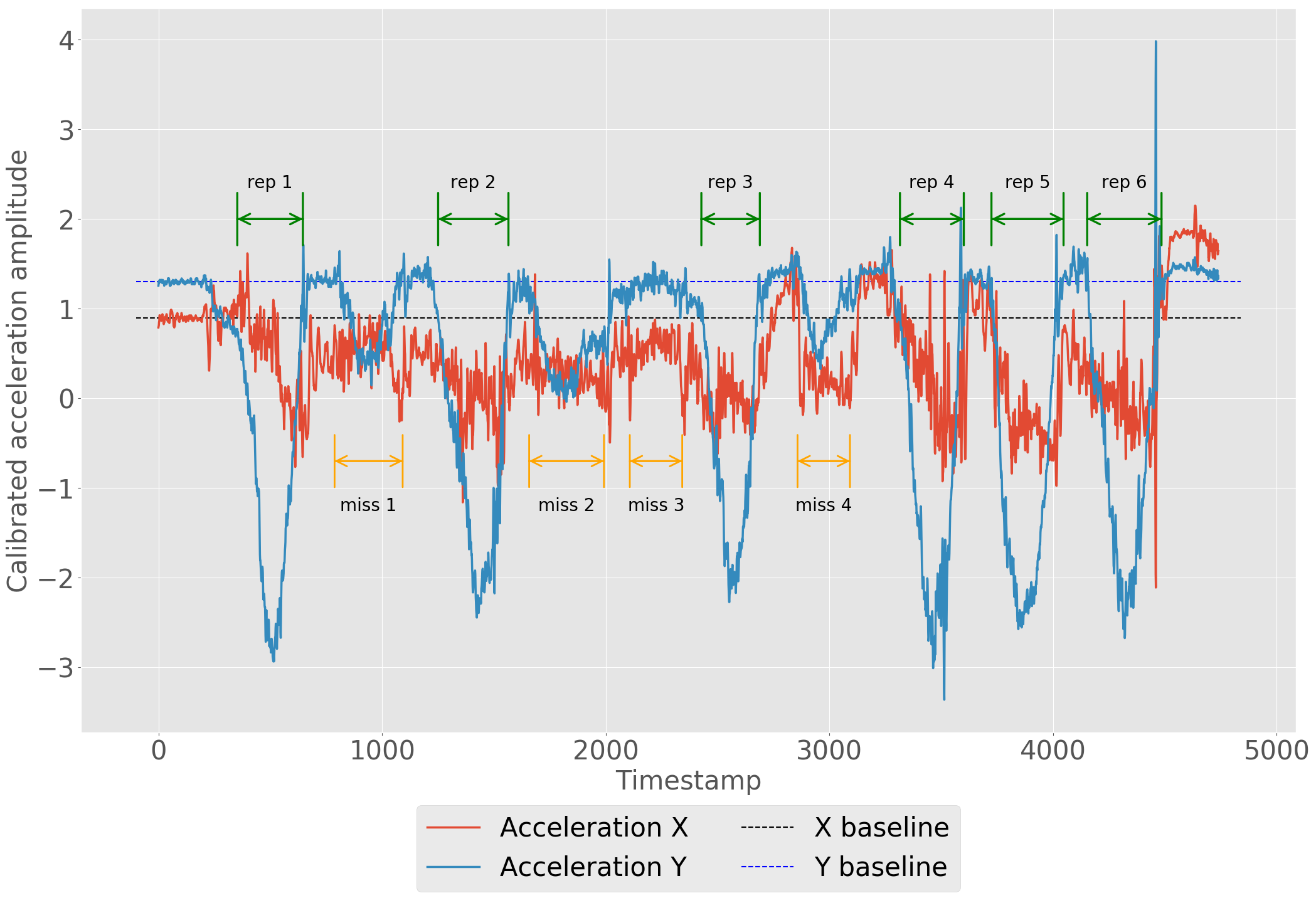}
  \caption{An instance of incorrectly segmented IRQ exercise.}
  \label{img:badSegmentedExercise}
\end{figure}

The proposed segmentation algorithm was able to properly detect all the
repetitions for HS and SAKE in the clinical datasets. All the incorrectly
chunked segments involve bad repetitions, both in mixed exercises and in wrong
exercises. An example is shown in Figure \ref{img:badSegmentedExercise}, where 4
repetitions out of 10 executed for an IRQ exercise performed by one of the
healthy subjects are not recognised. The first 4 repetitions of this particular
execution instance alternate between good and bad, then a bad-good-bad sequence
follows, then 3 good repetitions close the exercise. The exasperation of
simulated bad exercise executions led to poor segmentation results.

The classification results are listed in Table \ref{table:clsScores}. For each
exercise, the best values for accuracy, precision and recall are highlighted in
green. The SMO model was the most effective for SKE and SLR, while the Random
Forest performed best on HS and IRQ. Due to its narrow movement range, IRQ
obtained the less satisfying results, especially in the laboratory dataset.

\section{Conclusions and future work}
\label{section:futureWork}

In this paper, we presented a system that provides an automatic feedback to
patients on knee rehabilitation exercises. The collected data come from a
heterogeneous group of subjects, under a wide range of conditions. The
experimental results obtained from our campaigns show that the system is very
promising, thus further work is recommended. A usability study should validate
the performance of our Android application when adopted by real clinical
patients, as well as the accuracy level obtained for the classification of full
rehabilitation exercises, for which a minimum number of correct repetitions is
required. Different segmentation techniques should be tested, and the quality of
segments obtained from online algorithms should be properly assessed. In this
regard, Fourier transformation of the signals or interpolating functions plays a
key role in our future investigation. An exploratory study can also be performed
by using data collected with different device setups or by using different
hardware. In particular, the capabilities of common smartphone sensors hold a
significant research interest, as they often present a lower sampling frequency
when compared with medical devices.


\end{document}